\documentstyle[aps,preprint]{revtex}
\begin {document}

\draft
\title{Statics, metastable states and barriers in protein folding:\\
       A replica variational approach.}
\author{Shoji Takada and Peter G. Wolynes}
\address{School of Chemical Sciences, University of Illinois, Urbana, IL, 61801}
\date{}
\maketitle

\widetext
\begin{abstract}
Protein folding is analyzed using a replica variational 
formalism to investigate some free energy landscape characteristics relevant for dynamics. 
A random contact interaction model that satisfies the minimum frustration principle is
 used to describe the coil-globule transition (characterized by $T_{\rm CG}$), 
glass transitions (by $T_{\rm A}$ and $T_{\rm K}$) and folding transition (by $T_{\rm F}$).
Trapping on the free energy landscape is characterized by 
two characteristic temperatures, one dynamic, $T_{\rm A}$ the other static, $T_{\rm K}$ 
($T_{\rm A} > T_{\rm K}$), which are similar to those found in mean field theories of the
Potts glass. 1)Above $T_{\rm A}$, the free energy landscape is monotonous and 
polymer is melted both dynamically and statically. 2)Between $T_{\rm A}$ and $T_{\rm K}$, 
the melted phase 
 is still dominant thermodynamically, but frozen metastable states, 
exponentially large in number, appear. 3)A few lowest minima become thermodynamically 
dominant below $T_{\rm K}$, 
where the polymer is totally frozen.
In the temperature range between $T_{\rm A}$ and $T_{\rm K}$, 
barriers between metastable states are shown 
to grow with decreasing temperature suggesting super-Arrhenius behavior in a sufficiently 
large system. Due to evolutionary constraints on fast folding, 
the folding temperature $T_{\rm F}$ is expected to be
higher than $T_{\rm K}$, but may or may not be higher than $T_{\rm A}$.
Diverse scenarios of the folding kinetics are discussed 
based on phase diagrams that take into account the dynamical transition, 
as well as the static ones.
\end{abstract}


\narrowtext

\section{INTRODUCTION}
\label{sec:intro}
\par\noindent
In recent years the problem of protein folding, how a biological molecule 
spontaneously organizes itself under appropriate thermodynamic conditions, 
has become a fertile field of investigation for statistical physics
\cite{Bryngelson95,Chan93,Wolynes91,GarelXX}. 
The conceptual difficulty of finding the global free energy minimum, or 
native structure, reliably in a short amount of time, the so-called 
{\em Levinthal's paradox}\cite{Levinthal69} has come to be understood as being related to 
the problem of broken ergodicity in glassy systems\cite{Mezard87}.
In the modern version of the paradox, however, it is not the size of the 
configurational search alone that is relevant but rather the topography of 
the free energy landscape. The size of the free energy barriers between the 
metastable states of finite size heteropolymer determines the local rate of 
exploration of the free energy landscape. In addition, the global topography, 
in particular, whether there is an energetic bias funneling\cite{Leopold92} the molecule 
forwards a native structure is also important to understand the folding rate.

The earliest analytical approach to the problem captured these two aspects of the 
problem - the multiple minima problem and the guiding forces with the simplest 
description of the free energy landscape\cite{Bryngelson87,Bryngelson89}. 
The ruggedness of the free energy surface 
was modeled by the random energy model (REM)\cite{Derrida81}. The REM is the simplest model of a 
system which like a spin glass is frustrated through the conflict of 
many competing randomly chosen interactions.
A sufficiently large system with this free energy landscape was shown to possess 
a Levinthal paradox in its folding.
More precisely, at a characteristic glass transition temperature $T_{\rm K}$, 
while the system may thermodynamically prefer to be in unique configuration, 
the time to search for it would scale exponentially in the system size. 
(`K' of $T_{\rm K}$ is in honor of Kauzmann who attracted notice to the entropy crisis 
as the origin of the glass transition\cite{Kauzmann48}. See below for more details.)
Proteins are finite, however, so it is a quantitative issue whether such a 
system can fold on relevant biological time scales.
Buttressed by this asymptotic argument, but also calling upon observed regularities 
in protein structure, Bryngelson and Wolynes argued that most proteins are 
not random but additionally satisfy a {\rm principle of minimal frustration}, 
so that conflicts in attempting to satisfy individual interactions, 
are less than expected allowing a transition to a 
unique configuration at a folding temperature $T_{\rm F}$ higher than $T_{\rm K}$. 
The coherent part of the interactions could be taken account in the statics 
by introducing a conventional order parameter for folding, as in mean field 
theory. For a small system this order parameter can also act as an approximate global reaction 
coordinate for describing the self-organization process\cite{Bryngelson89}.
This relatively simple framework can be elaborated to take into account 
additional order parameters for folding, such as local secondary 
structure formation\cite{Saven96} and recently correlations in the free energy 
landscape\cite{Plotkin96}.
The framework and the resulting mechanistic scenarios are also quite useful 
for organizing the discussions of many experiments\cite{Bryngelson95}.

Another significant thread in the statistical physics of protein folding has 
been provided by theories that use the replica technology of spin glass theory\cite{Mezard87} 
along with polymer physics to understand the free energy landscape
\cite{Garel88,Shakhnovich89,Sasai90,Ramanathan94,Pande94,Hao95}. 
Garel and Orland\cite{Garel88}, as well as Shakhnovich and Gutin\cite{Shakhnovich89} 
studied random heteropolymers using 
the traditional polymeric virial expansion Hamiltonian of a connected chain 
incorporating a Gaussian random pair interaction. These workers showed the connection 
of the random heteropolymer thermodynamics with the phase transition of Potts glass
\cite{Gross85}.
Qualitatively this was not entirely unexpected  because a wide range of frustrated 
random systems without special symmetries fall in this universality class, 
which also includes the REM model\cite{Gross85,Kirkpatrick89,Kirkpatrick87a,Kirkpatrick87b}.
This work was relevant to the ruggedness issues but not to the problem of 
guiding forces. 
Soon after this work, Sasai and Wolynes dealt with three aspects, polymeric interaction, 
ruggedness of free energy landscape, and results of evolution, 
 in one model\cite{Sasai90}. They employed a variational approach modeled on 
Feynman's polaron theory\cite{FeynmanPO} in the replica 
space to analyze the associative memory Hamiltonian\cite{Friedrichs89}. 
1)This model has explicit chain 
connectivity. 2)The target structure, i.e., the native structure, 
included in the memory set (or data base) provides a route to incorporate 
the role of principle of minimal frustration, while 
3)memories other than the target induce ruggedness in the free energy landscape.
Very recently, Ramanathan and Shakhnovich shed light on effects of the evolutionary constraint
of minimal frustration in more detail\cite{Ramanathan94}. 
Instead of assuming the pronounced energy gap {\it a priori}, they represented
evolution as a process that yields sequences distributed according to a Boltzmann 
distribution for a 
fixed target structure. Their theory shows that it is possible to have an
energy gap large enough to stabilize the native structure only by choosing the
sequence appropriately although it is not clear 
if nature actually used such a sequence selection mechanism or not.
An alternative route to minimal frustration called `imprinting' has also been 
discussed by Pande et al.\cite{Pande94}, which finally gives almost same result as 
\cite{Ramanathan94}.

Levinthal's paradox makes stark that the conceptual issues of the folding problem 
revolve on {\em kinetics} in at least a semi-quantitative fashion. 
Theory and many simulations\cite{Socci94,Sali94} in concurrence suggest that real
proteins fold below their folding temperature $T_{\rm F}$ but somewhat above 
the (static) glass phase transition temperature $T_{\rm K}$. 
Thus, to understand the kinetics of folding, a microscopic description of the free energy 
landscape above $T_{\rm K}$ is indispensable. 
Is the effective free energy landscape monotonous and smooth above $T_{\rm K}$? 
We claim no. Even above $T_{\rm K}$ there are a number of local minima 
lasting many vibrational periods (Rouse relaxation times) in the 
free energy landscape. Although the variational solution corresponding with the melt phase
 dominates the formal Boltzmann 
average, actually a protein is dynamically trapped and feels some of the 
ruggedness of the free energy landscape 
and thus kinetics would strongly be 
affected by the presence of local minima. 
Then the next question that arises concerns the barrier heights between 
these local minima because these barrier heights determine the kinetics. 
We show that barrier heights grow with decreasing temperature until $T_{\rm K}$ is reached, 
which directly leads us to the super-Arrhenius activation behavior in this 
temperature regime.

To make this analysis we utilize recently developed ideas in the spin glass theory, 
especially for the Potts-type spin glass
\cite{Kirkpatrick87a,Kirkpatrick87b,Crisanti92,Kurchan93}.   
In a series of papers Kirkpatrick, Thirumalai, and Wolynes, working on 
models of structural glasses\cite{Kirkpatrick89}, $p-$spin interaction model 
glasses($p>2$)\cite{Kirkpatrick87a}
and the Potts glasses with more than 4 components\cite{Kirkpatrick87b}  
made the following observations:
1)The phase transition temperature $T_{\rm A}$ obtained by the dynamical theory, 
i.e., mode-mode coupling theory based on Langevin dynamics, is higher 
than that $T_{\rm K}$ obtained by the static theory, i.e., the ordinary replica method. 
2)As temperature decreases starting from the paramagnetic phase, solutions of 
the Thouless-Anderson-Palmer (TAP) equations\cite{Thouless77} except paramagnetic 
one appear exactly at $T_{\rm A}$ (See Fig.\ref{fg:TAP}). 
3)For $T_{\rm A}>T>T_{\rm K}$, many metastable states are 
separated by high barriers and therefore have a long lifetime. Thus 
activated transport is the typical picture in this range (`A' of 
$T_{\rm A}$ means `activation'). 
3)The overlap order parameter, $q$, in the same group of 1 level
replica symmetry breaking (RSB) takes a discontinuous jump at $T_{\rm K}$, 
which reminds us of a first order phase transition in the order parameter, 
but the transition looks a second order in that thermodynamically, 
there is no latent heat. (This was known and well understood in the case 
of REM.) They called this class of phase transitions, {\em random first order 
phase transitions}. 
Crisanti and Sommers found essentially the same behavior in the $p$-spin spherical 
model\cite{Crisanti92}, which buttresses the case that this type of behavior, 
very different from that of the Sherrington-Kirkpatrick model, is quite universal for 
systems without inversion symmetry. Using the $p$-spin spherical model 
Kurchan, Parisi and Virasoro succeeded in 
describing the metastable states in greater detail and the barriers between them 
in the replica formalism\cite{Kurchan93},
which we use in this paper. 
This formalism for describing metastable states has some forbidding aspects. 
Like the equilibrium replica technique there are steps involving analytical continuation
to apparently nonphysical values of replica number. More work to clarify the 
techniques would be welcome but the physical content seems very much in 
keeping with a transition is driven by configurational entropy. 
Barrier heights are determined by a competition between the number of available 
states and the energetic advantage which a polymer can achieve in a particular 
lower minimum. The results on barrier heights are the main focus of this paper. 

In this paper we employ the contact interaction model used in \cite{Shakhnovich89} 
with the principle of minimal frustration implemented at the level of \cite{Sasai90}. 
Methodologically, we 
rely on the replica variational approach of Sasai and Wolynes, 
but extend the interpretation to the level of Kurchan, Parisi, and Virasoro\cite{Kurchan93}
 for metastable states and barriers.
These methods are summarized in section II.
In section III, we introduce some approximations so that we can derive expressions for the 
free energy in as simple form as possible.  
These expressions are used in section IV to locate the phase 
transitions between different phases. We derive explicit expressions for four 
phase transition temperatures, the coil-globule transition temperature $T_{\rm CG}$, 
the folding temperature $T_{\rm F}$, the dynamical glass temperature $T_{\rm A}$, 
and the static glass temperature $T_{\rm K}$. 
In particular, the ruggedness of free energy landscape is 
characterized by two critical temperatures of freezing, $T_{\rm A}$ and $T_{\rm K}$ 
as is in the case of Potts glass. 
In section V we draw phase diagrams with fairly diverse states and discuss several scenarios 
of the folding kinetics, which can be thought as a refined version of the scenarios 
given in \cite{Bryngelson95}.
Complete but somewhat messy expressions for the free energy are given
in the appendix.

\section{REPLICA VARIATIONAL APPROACH}
\label{sec:replica}
\subsection{Model}
The model we present here, while different from that used by Sasai and Wolynes\cite{Sasai90}, 
is motivated by it. Our main goal in this section is to show how a model with 
short range (in space) interaction can be treated with the same formalism as the 
long range associative memory model.

As a simple model of protein, we start with a standard beads type Hamiltonian 
which includes the interaction between monomers in the form of the virial expansion.
\begin{eqnarray}\label{eq:Horg}
H&=& k_BT \sum_{i} {({\bf r}_{i+1}-{\bf r}_{i})^2\over 2a^2}
    +{v\over 2} \sum_{i\neq j}b_{ij}\delta ({\bf r}_{i}-{\bf r}_{j})
    +c{v^2\over 6} \sum_{i\neq j\neq k}\delta ({\bf r}_{i}-{\bf r}_{j})
                                           \delta ({\bf r}_{j}-{\bf r}_{k}),
\end{eqnarray}
where ${\bf r}_i$ represents $\alpha$ carbon of each amino acid ($i=1\sim N$), 
$a$ is the Kuhn length\cite{Kuhn}, $v$ represents finite resolution of 
space (see below), $b_{ij}$ and $c$ 
are the second and third virial coefficients, respectively. Depending on the 
type of amino acids individual $b_{ij}$ have apparently random values, whose
 distribution will be given below.
We assume the spatial resolution is $v^{1/3}$ and so any function is 
smeared out within this scale. Therefore, $\delta ({\bf 0})=v^{-1}$.
The above Hamiltonian itself is directly suitable to the random 
heteropolymer, as was used in \cite{Shakhnovich89}.

Since a protein can fold because of its specific sequence, it is indispensable 
to incorporate the principle of minimal frustration, as was mentioned in the Introduction.
The key idea here is 
that the energy of ground state, which corresponds to the target structure defined by amino 
acid positions $\{{\bf r}_i^{\rm T}\}$ of the native state,
depends strongly on the specific sequence of amino acids, while properties of 
non-native structures can well be modeled by the random interaction between 
amino acids. In other words, the energy of native structure is non-self averaging, 
while most others that are structurally unrelated are self averaging.
This is supported by numerical enumeration of all the compact states in 
the lattice 27-mer\cite{Sali94}. 
Using a measure of nativeness,
\begin{equation}
q={v\over N}\sum_i \delta \left({\bf r}_i-{\bf r}_i^{\rm T}\right),
\end{equation}
we rewrite the above Hamiltonian separating the non-self-averaging part 
from the others,
\begin{eqnarray}\label{eq:H}
H&=& k_BT \sum_{i} {({\bf r}_{i+1}-{\bf r}_{i})^2\over 2a^2}
    +(1-q){v\over 2} \sum_{i\neq j}b_{ij}\delta ({\bf r}_{i}-{\bf r}_{j}) \nonumber\\
 &+&(1-q)c{v^2\over 6} \sum_{i\neq j\neq k}\delta ({\bf r}_{i}-{\bf r}_{j})
                                           \delta ({\bf r}_{j}-{\bf r}_{k})
      +qE^{\rm T},
\end{eqnarray}
where
\begin{eqnarray}
qE^{\rm T}&=& {v\over N}\sum_i \delta \left({\bf r}_i-{\bf r}_i^{\rm T}\right)\left[
       {v\over 2} \sum_{i\neq j}b_{ij}\delta ({\bf r}_{i}-{\bf r}_{j}) 
 +c{v^2\over 6} \sum_{i\neq j\neq k}\delta ({\bf r}_{i}-{\bf r}_{j})
                 \delta ({\bf r}_{j}-{\bf r}_{k})\right].
\end{eqnarray}
Here we introduce an approximation,
\begin{eqnarray}\label{eq:ET}
qE^{\rm T} &\simeq& {v\over N}\sum_i \delta \left({\bf r}_i-{\bf r}_i^{\rm T}\right)\left[{v\over 2} \sum_{i\neq j}b_{ij}\delta ({\bf r}_{i}^{\rm T}-{\bf r}_{j}^{\rm T}) 
 +c{v^2\over 6} \sum_{i\neq j\neq k}\delta ({\bf r}_{i}^{\rm T}-{\bf r}_{j}^{\rm T})
                \delta ({\bf r}_{j}^{\rm T}-{\bf r}_{k}^{\rm T})\right],
\end{eqnarray}
which is exact either when the system is in the native structure or when the system 
is totally uncorrelated to the native structure. 
In eq.(\ref{eq:H}), the second and third terms are assumed to self-averaging, 
while the last term is non-self-averaging.
{\em After} the non-self-averaging term representing minimal frustration of the
target structure is taken into account,  
the interaction energies $b_{ij}$ in eq.(\ref{eq:H}) may be modeled, as was mentioned,
by Gaussian random variables with probability distribution,
\begin{equation}\label{eq:bij}
P\left( b_{ij}\right) = (2\pi b^2)^{-1/2} \exp \left[ -(b_{ij}-b_0)^2/ 2b^2 \right].
\end{equation}
Note that we do {\em not} take an average of $b_{ij}$ in eq.(\ref{eq:ET}), 
which are thought as sequence specific.
Equations (\ref{eq:H}), (\ref{eq:ET}), and (\ref{eq:bij}) defines the model, in which 
parameters, $T$, $b_0$, $b$, and $E^{\rm T}$ play central roles. 

Here, we should bear in mind that 
the virial expansion is, as is well-known, good for extended states such as the random 
coil state but not very accurate for highly collapsed state, which we are mainly interested in.
Thus, the present thermodynamic description of the radius of polymer, 
in particular, may not be particularly accurate. 

\subsection{Replica variational formalism and mean field approximation}
We summarize the variational polaron approach in replica space used earlier\cite{Sasai90}.
We calculate the free energy $[F]_{\rm av}=-k_BT[\ln Z]_{\rm av}$ averaged over the random 
bond interaction $b_{ij}$ with probability distribution eq.(\ref{eq:bij}), 
where $[\quad ]_{\rm av}$ means the average over $b_{ij}$ and 
$Z$ is the canonical partition function.
To avoid the difficulty of taking an average of $\ln Z$,  
the replica trick\cite{Mezard87} utilizes a mathematical identity, 
$\ln x=\lim_{n\rightarrow 0}(x^n-1)/n$. Thus, 
\begin{equation}
-\beta \left[ F \right]_{\rm av} = \left[ \ln Z \right]_{\rm av} = \lim_{n\rightarrow 0}
{\left[ Z^n \right]_{\rm  av}-1 \over n}.
\end{equation}
We then concentrate on $[Z^n]_{\rm av}$, which is explicitly given as
\begin{eqnarray}\label{eq:Znav}
\left[ Z^n \right]_{\rm av} &=& \int \prod_{i>j}\left[{\rm d}b_{ij} P\left( b_{ij}\right)\right]
\int \prod_{\alpha=1}^{n} {\cal D}{\bf r}_i^\alpha 
{\rm e}^{-\beta \sum_\alpha H\left(\{ {\bf r}_i^\alpha \}\right) },
\end{eqnarray}
where 
\begin{equation}
{\cal D}{\bf r}_i \equiv \prod_i d{\bf r}_i \delta \left(\sum_i {\bf r}_i\right).
\end{equation}
The delta function in the above equation is used to fix the center of mass at 
the origin. Since the integrand in eq.(\ref{eq:Znav}) is a Gaussian function with 
respect to $b_{ij}$, we can integrate $b_{ij}$ out at the beginning to get
\begin{eqnarray}\label{eq:Zn}
\left[ Z^n \right]_{\rm av} &=& 
\int \prod_{\alpha} {\cal D}{\bf r}_i^\alpha 
{\rm e}^{-\beta H_{\rm eff}}.
\end{eqnarray}
The effective Hamiltonian here is of the form 
\begin{equation}
H_{\rm eff}=H_0+H_1+H_2,
\end{equation}
each term of which is given as
\begin{equation}
H_0=k_BT \sum_{\alpha ,i} {({\bf r}_{i+1}^\alpha-{\bf r}_{i}^\alpha)^2\over 2a^2},
\end{equation}
\begin{eqnarray}
H_1&=& \sum_\alpha q_\alpha E^{\rm T} +\sum_{\alpha}{v\over 2}
\left[ b_0 (1-q_\alpha)-{\beta b^2\over 2}(1-q_\alpha)^2 \right]
\sum_{i\neq j}\delta ({\bf r}_{i}^\alpha-{\bf r}_{j}^\alpha)\nonumber\\
  &+&c{v^2\over 6}\sum_\alpha (1-q_\alpha) 
\sum_{i\neq j\neq k}\delta ({\bf r}_{i}^\alpha-{\bf r}_{j}^\alpha)
                    \delta ({\bf r}_{j}^\alpha-{\bf r}_{k}^\alpha)
\end{eqnarray}
and
\begin{eqnarray}
H_2&=&-{\beta b^2 v^2 \over 4} \sum_{\alpha\neq\beta}(1-q_\alpha)(1-q_\beta)
\sum_{i\neq j}\delta ({\bf r}_{i}^\alpha-{\bf r}_{j}^\alpha)
              \delta ({\bf r}_{i}^\beta -{\bf r}_{j}^\beta ).
\end{eqnarray}
$H_0$ maintains the polymeric chain connectivity, $H_1$ includes the one-replica 
part, and $H_2$ represents the inter-replica interaction. Obviously, the latter 
term is the driving force of the RSB.

Integration over the vast configuration space in eq.(\ref{eq:Zn}) is too complicated 
to execute exactly. So, we generalize the variational principle well-known in 
the statistical physics\cite{Feynman65} to replica space; For any 
reference Hamiltonian $H_{\rm ref}\left(\{{\bf r}_i^\alpha\}\right)$, we have an 
inequality relation,
\begin{equation}\label{eq:var-ineq}
F_{\rm var}\equiv F_{\rm ref}+\langle H_{\rm eff}-H_{\rm ref}\rangle \geq F_{\rm eff},
\end{equation}
where 
$$
-\beta F_{\rm ref}=\ln Z_{\rm ref}=\ln \int \prod_{\alpha} {\cal D}{\bf r}_i^\alpha 
{\rm e}^{-\beta H_{\rm ref}},
$$
$-\beta F_{\rm eff}=\ln [Z^n]_{\rm  av}$ and $\langle \cdots \rangle$ means 
an expectation value for the Hamiltonian $H_{\rm ref}$. 
This inequality holds before we take a limit $n\rightarrow 0$.
Using this principle we optimize $F_{\rm var}$ with respect to order parameters 
included in the reference Hamiltonian. With the optimized $F_{\rm var}^*$, we get 
an estimate of the free energy we are seeking,
$[F]_{\rm av}=\lim_{n\rightarrow 0}F_{\rm eff}/n\simeq \lim_{n\rightarrow 0}F_{\rm var}^*/n$.

Reference trial functions need to be simple enough to lead to soluble partition 
function but flexible enough to include order 
parameters which characterizes all relevant phase transitions. 
The coil-globule transition, the folding transition, the glass transition,
 can be characterized by the radius of gyration, by a fluctuation scale around the 
native structure, and by an inter-replica correlation (Debye-Waller 
factor in the glass phase), respectively. A natural choice for a reference 
Hamiltonian is 
\begin{eqnarray}
\beta H_{\rm ref}&=&A \sum_{\alpha ,i}({\bf r}_{i+1}^\alpha-{\bf r}_{i}^\alpha)^2
  +B \sum_{\alpha ,i} ({\bf r}_i^\alpha)^2
  +C \sum_{\alpha ,i} ({\bf r}_{i}^\alpha-{\bf r}_{i}^{\rm T})^2
  +D \sum_{\alpha\neq\beta ,i} d_{\alpha\beta}({\bf r}_{i}^\alpha-{\bf r}_{i}^{\beta})^2,
\end{eqnarray}
where $A=(2a^2)^{-1}$ and all $B$, $C$, $D$, and $d_{\alpha\beta}$ are 
free parameters to be optimized based on the variational principle eq.(\ref{eq:var-ineq}).
Once these parameters are optimized, they play the role of the global order parameters;
$B$, $C$, $D$, and $d_{\alpha\beta}$ represents the radius of gyration, 
fluctuation around the native structure, inter-replica correlation, and 
the mode of RSB, respectively. 

As for the mode of RSB, we rely on analogy to the Potts glass with components more 
than 4. As we mentioned, many other models exhibit the same type of RSB and this 
is believed to be quite universal for the system without inversion symmetry. 
In this class of systems, one level 
of Parisi's RSB scheme has been shown to be sufficient to describe the stable and 
metastable states\cite{Kurchan93} and we concentrate on this level of description in this paper. 
Then, $n$ replicas are divided into $n/m$ groups, each of which has size $m$ 
and the matrix element $d_{\alpha\beta}$ is 1 if $\alpha$ and $\beta$ ($\alpha\neq\beta$) 
belong to the same group and 0 otherwise.

\subsection{Free energy}

We just give an overview of the derivation and the 
 expression for the variational free energy $F_{\rm var}$ 
defined in eq.(\ref{eq:var-ineq}) here. Detailed expressions may be found in the Appendix 
for completeness since these are not important to understand the present arguments.  
Physically, the free energy $F_{\rm var}$ consists of three parts, a conformational 
entropy term $-nTS$, a one-replica part $\langle H_1 \rangle$ which contains 
the coherent part of the interactions, which ultimately give a folding funnel 
as well as an effective homopolymer term, 
 and the inter-replica term $\langle H_2\rangle$ which is responsible for the random 
interaction between monomers. We explain each of them.

In order to carry out the variational procedure 
we start with the calculation of $Z_{\rm ref}$, which is the same as that of 
Sasai and Wolynes\cite{Sasai90}. More details can be found in \cite{Sasai90}.
We first diagonalize $d_{\alpha\beta}$ with respect to the replica index. 
Concentrating on each block of size $m$, we get two type of eigenmodes, 
a symmetric mode with the eigenvalue $\Lambda_+=0$ (we call $+$ mode) and 
$m-1$ degenerate asymmetric modes with the eigenvalue $\Lambda_-=2m$ ($-$ mode).
With the diagonalized replica index $\mu$, we 
see that the integrand is just a Gaussian function of ${\bf r}_i^\mu$. 
The exponent is of the form $A\sum_{\mu ij} {\bf r}_i^\mu {\cal H}_{ij}^\mu {\bf r}_j^\mu$, 
where the coefficient matrix ${\cal H}^\mu$ ($\mu=\pm$) is 
$$
{\cal H}^\pm =\left(\begin{array}{ccccccc}
2\cosh \lambda_\pm-1  & -1 & 0  & 0 & \cdots & \cdot & 0       \\
-1 & 2\cosh\lambda_\pm     & -1 & 0 & \cdots & \cdot & 0       \\
0  & -1 & 2\cosh\lambda_\pm    & -1 & \cdots & \cdot & 0       \\
\cdot & \cdot & \cdot & \cdot       & \cdots & \cdot & \cdot   \\
&  &  & \cdots & -1   & 2\cosh\lambda_\pm            & -1      \\
&  &  & \cdots & 0 & -1     & 2\cosh\lambda_\pm-1       \end{array}
\right),
$$
where $\lambda_\pm$ is defined by 
$2\cosh\lambda_\pm=2+(B+C+\Lambda_\pm D)/A.$
Thus, it is straightforward, although complicated, to integrate over 
configuration space and the result is represented in terms of $G_{ij}^\pm$, the 
inverse matrix of ${\cal H}^\pm$. $F_{\rm ref}$ is given as a function of 
$B$, $C$, $D$, and $m$, explicit formula of which is given in the 
Appendix.

The conformational entropy $S$ is expressed by
\begin{eqnarray}
nTS&=& - F_{\rm ref} + \langle H_{\rm ref} \rangle 
      - \langle H_0 \rangle,
\end{eqnarray}
First, $F_{\rm ref}$ is simply obtained as $\beta F_{\rm ref}=-\ln Z_{\rm ref}$.
Second, $\langle H_{\rm ref} \rangle$ can be evaluated from the scaling 
argument. If we scale as ${\bf r}_i\rightarrow \sqrt{z}{\bf r}_i^\prime$, the 
exponent of the integrand changes $\beta H_{\rm ref}({\bf r}_i)\rightarrow 
z\beta H_{\rm ref}({\bf r}_i^\prime)$ because $H_{\rm ref}$ is a homogeneous 
quadratic function of ${\bf r}_i$. Thus, taking a derivative of $\ln Z_{\rm ref}$
 written in terms of ${\bf r}_i^\prime$ with respect to $z$ we get an expression
 for $\langle H_{\rm ref} \rangle$. 
Finally, $\langle H_{0} \rangle$ is simply given by $\langle H_{0} \rangle=-k_BTA(\partial 
\ln Z_{\rm ref}/\partial A)$. 
The conformational entropy $S$ expressed thus in terms of order parameters $B$, $C$, 
$D$, and $m$ is explicitly written in the Appendix.

For the estimate of $\langle H_1 \rangle$, we introduce an additional approximation 
in the spirit of the mean field theory. 
Defining the monomer density $\rho_\alpha ({\bf r})$ as
$\rho_\alpha ({\bf r})\equiv\sum_i \delta ({\bf r}-{\bf r}_i^\alpha)$, 
replacing expectation value of products by the products of 
expectation values (mean-field approximation), we get
\begin{eqnarray}
\langle H_1 \rangle&\simeq& \sum_\alpha \langle q_\alpha \rangle E^{\rm T} 
                    +\sum_{\alpha }{v\over 2}\left[ b_0 (1-\langle q_\alpha\rangle)
                          -{\beta b^2\over 2}(1-\langle q_\alpha\rangle)^2 \right]
       \int \langle\rho_\alpha ({\bf r})\rangle^2 d{\bf r}\nonumber\\
  &+&c{v^2\over 6}\sum_\alpha (1-\langle q_\alpha\rangle) 
\int \langle \rho_\alpha ({\bf r})\rangle^3 d{\bf r}.
\end{eqnarray}

In the same way, introducing the overlap order parameter function 
$$
Q_{\alpha\beta}({\bf r}_1,{\bf r}_2)
=\sum_i \delta ({\bf r}_1-{\bf r}_i^\alpha) 
        \delta ({\bf r}_2-{\bf r}_i^\beta),
$$ 
we can express $H_2$ as
\begin{eqnarray}
\langle H_2\rangle&\simeq&-{\beta b^2 v^2 \over 4} 
\sum_{\alpha\neq\beta}(1-\langle q_\alpha\rangle)(1-\langle q_\beta\rangle)
\int\int d{\bf r}_1 d{\bf r}_2 \langle Q_{\alpha\beta}({\bf r}_1,{\bf r}_2)\rangle^2.
\end{eqnarray}
$\langle\rho_\alpha\rangle$, $\langle q\rangle$, and $\langle Q_{\alpha\beta}\rangle$ 
can be calculated by direct integration and are expressed in terms of 
$B$, $C$, $D$, and $m$, which are given in the Appendix. 
We note that it is possible to calculate $<H_1>$ and $<H_2>$ without introducing 
these approximations; the result becomes more complex but does not change the 
argument discussed in this paper. Therefore, we employ this approximation 
to get simpler expressions keeping the qualitative results unchanged. 
It is also advantageous to use these approximations in that it makes it easy 
to compare our results with those of Shakhnovich and Gutin\cite{Shakhnovich89}.

\section{COIL, GLOBULE, GLASS AND FOLDED PHASES}
\label{sec:F}

Although the above results are quite general, it is hard to grasp the 
physical picture directly from them without any numerical work. 
Therefore, we introduce several other approximations to get simple analytical 
expressions for free energy. 
We take a sort of self-consistent strategy in the following way. 
First, we assume for each phase that one specific 
order parameter (or $A$) is much larger than the others. Second, using 
this inequality, we obtain an asymptotic expression for the free energy 
and seek the stationary solution with respect to order parameters for each phase. 
Finally, we confirm that the solution indeed satisfies the inequality we assumed.  

The first approximation introduced is that $N \gg 1$ and most of non-extensive 
terms are ignored. This may actually be a severe approximation for practical 
work since proteins are mesoscopic and possess a considerable surface area. 
Next, we employ the simplest description of monomer 
density, the so-called volume approximation\cite{Grosberg94,volume}; 
$\langle\rho ({\bf r})\rangle$ is a positive constant $\rho$ inside polymer 
and is zero outside. Thus, $\int \rho^x({\bf r}) d{\bf r}=V\rho^x=N\rho^{x-1}$, where 
$V$ is the total volume of the polymer and $x$ is an integer.
Thirdly, we 
approximate $\int <Q_{\alpha\beta}>^2 d{\bf r}_1 d{\bf r}_2$ as (see Appendix)
$$
\int Q_{\alpha\beta}^2 d{\bf r}_1 d{\bf r}_2\simeq 
N\rho \left({4\pi G_{ii}^-\over A} \right)^{-3/2},
$$
(Hereafter, we drop $\langle\cdots\rangle$, for simplicity),
for the case $\alpha$ and $\beta$ belonging to the same group of 1 level RSB. 
For the other cases, $G_{ii}^-$ is replaced by $g_i\equiv {1\over m}\left[ G_{ii}^+ 
+(m-1)G_{ii}^- \right]$. 
The other approximations 
we use are dependent on the phase we consider and will be explained below 
one by one. 

\subsection{Coil and globule phases}

The coil and globule phases may be characterized by the inequality $A \gg B, C, 2mD$.
Assuming this inequality, we consider the radius of gyration defined by
\begin{equation}
R=\sqrt{\langle \sum_{\alpha,i}{\bf r}_i^{\alpha 2}\rangle/nN},
\end{equation}
and we get an asymptotic expression,
$$
R^2 = {3\over 4}\left[{1\over m}{1\over\sqrt{A(B+C)}}
              +{m-1\over m}{1\over\sqrt{A(B+C+2mD)}}\right].
$$
Random-coil state should have the radius $R\sim N^{1/2}a$. Combining it with 
the above equation we see that $B+C$ and $2mD$ are at most of order $N^{-2}A$,
 which is consistent 
with the inequality we assumed. 
In the same way, the radius scales as $R\sim N^{1/3}a$ in the globule phase, 
which leads us to the estimate $B+C\sim N^{-4/3}A$ and $2mD\sim N^{-4/3}A$.
Approximating that the polymer is roughly spherical with radius $R$, 
$\rho$ can be related to $R$ by
$$
\rho={N\over (4/3) \pi R^3}.
$$
An estimate of the free energy is quite straightforward. 
First, starting from the full expression given in the Appendix, 
we can derive an asymptotic expression for the entropic part,
$$
-TS={3\over 4}Nk_BT\left[ {1\over m}\sqrt{B+C\over A}
                         +{m-1\over m}\sqrt{B+C+2mD\over A}\right],
$$
which is of order $O(N^0)$ for the coil state and is $O(N^{1/3})$ for the globule 
state.
Secondly, the inter-replica term $H_2$ is of order $O(N^{-1/2})$ for the 
coil state and $O(N^0)$ for the globule state and thus is negligible.

For convenience, we change the independent variables from $B$, $C$ and $2mD$ to $\rho$, 
$q$, and $2mD$. Then, we can easily optimize $2mD$ and $q$ to get the 
solution $2mD=q=0$ (under some conditions discussed below), which leads to
$$
R^2 = {3\over 4}{1\over\sqrt{A(B+C)}}=\left( {3N\over 4\pi\rho}\right)^{2/3}.
$$
and  
$$
-TS={3\over 4}Nk_BT \sqrt{B+C\over A}\propto N^{1/3}k_BT \rho^{2/3}.
$$
The latter is of order $O(N^0)$ for the coil state and $O(N^{1/3})$ for the globule state.

Thus, the free energy can be represented only as a function of $\rho$ as 
\begin{equation}\label{eq:FCG}
F_{\rm CG}=N{v\over 2}\left(b_0-{\beta b^2\over 2}\right)\rho
          +Nc{v^2\over 6}\rho^2 
          +N^{1/3}p \rho^{2/3} k_BT/A,
\end{equation}
where $[\cdots]_{\rm av}$ in the LHS is dropped for simplicity and 
$p$ is a constant of order unity, the value 
itself $(9/16)(4\pi/3)^{2/3}$is not important for the present purpose.
The first two terms are of the form of virial expansion with coefficients 
$b_0-\beta b^2/2$ and $c$; the random interaction induces an effective attraction 
proportional to $1/T$. The third term comes from entropy loss due to packing.
Although the latter is non-extensive and is not important for many situations, 
we retain it because it will play important roles for some cases  
as will be explained below.

\subsection{Glass phase}

Since the glass phase is characterized by a small thermal fluctuations around 
individual minima, we 
assume $2mD \gg A \gg B, C$. 
Using this relation we can straightforwardly obtain the asymptotic expression for 
the entropic contribution to the free energy as\cite{Sasai90},
$$
-TS={m-1 \over m}Nk_BT\left[ \ln \left({2mD\over A}\right)^{3/2}
-{3\over 2}\right].
$$
This can be interpreted as a confinement entropy.

Next, let us consider the random interaction part $H_2$. 
$G_{ii}^-$ behaves as $A/(2mD)$ in the present limit and we have
$\int Q_{\alpha\beta}^2 d{\bf r}_1 d{\bf r}_2\simeq N\rho(4\pi)^{-3/2}(2mD)^{3/2},$
for the case $\alpha$ and $\beta$ belong to the same group of RSB and $\sim 0$ otherwise.
Here, we have to take care of the finiteness of spatial resolution as mentioned before.
The above estimate holds only when $|{\bf r}_1-{\bf r}_2|\sim G_{ii}^-/A$ is 
of order $v^{1/3}$ or larger. Otherwise, $Q_{\alpha\beta}$ should be replaced 
by the delta function with $\delta ({\bf 0})=v^{-1}$, which gives 
$\int Q_{\alpha\beta}^2 d{\bf r}_1 d{\bf r}_2\simeq N\rho v^{-1}.$
To make the expression continuous with respect to the order parameter $D$, we switch two 
expressions when both take the same value. In summary, 
$$
\int Q_{\alpha\beta}^2 d{\bf r}_1 d{\bf r}_2\simeq\left\{
           \begin{array}{ll} (4\pi)^{-3/2}N\rho(2mD)^{3/2} & \quad {\rm if}\quad
                                                         (2mD)^{3/2} \leq (4\pi)^{3/2}/v \\
                            N\rho v^{-1}                   & \quad {\rm  otherwise}.
                   \end{array}\right. 
$$

In the same way as above, we change independent variables from $B$, $C$, $D$, and $m$ 
to $\rho$, $q$, $2mD$, and $m$. We can show that $q=0$ is stable unless the stability 
gap is too large and thus we can write down the free energy expression,
\begin{eqnarray}\label{eq:Fglass}
F_{\rm Glass}&=&N{v\over 2}\left(b_0-{\beta b^2\over 2}\right)\rho
             +Nc{v^2\over 6}\rho^2
             -N{\beta b^2 v^2\over 4}(m-1)\left\{
                   \begin{array}{l} (4\pi)^{-3/2}\rho(2mD)^{3/2} \\
                                    \rho v^{-1} 
                   \end{array}\right. \nonumber\\
             &+&N{m-1 \over m}k_BT\left[ \ln \left({2mD\over A}\right)^{3/2}-{3\over 2}\right],
\end{eqnarray}
where upper (lower) terms is taken when $(2mD)^{3/2}$ is smaller (larger) than
$(4\pi)^{3/2}/v$. The first two terms are those of virial expansion as above,  
the third term represents inter-replica interaction and is the driving force
for the RSB, and the last term is obviously entropic.

\subsection{Folded phase}

The folded phase is characterized by a large $C$, i.e., small fluctuations 
around an ideal native structure and so we assume $C \gg A, B, 2mD$.
We can easily obtain an asymptotic expression for the entropic part, as 
was done in \cite{Sasai90};
$-TS\simeq (3/2)Nk_BT \ln {C\over A}.$
The inter-replica part $H_2$ has the replica-symmetric contribution,
$\int Q^2_{\alpha\beta} d{\bf r}_1 d{\bf r}_2=(2\pi)^{-3/2}N\rho C^{3/2}$
for any pair of $\alpha$ and $\beta$.
Nativeness $q$ in this limit is obtained as 
$q\simeq v\left({C\over \pi}\right)^{3/2}.$
Using this we change independent variables from $B$, $C$, and $D$ to 
$\rho$, $q$, and $2mD$.

We can show that $2mD=0$ is the stable solution and thus
\begin{eqnarray}\label{eq:Ffolded}
F_{\rm Folded}&=&N{v\over 2}\left[b_0(1-q)-{\beta b^2\over 2}(1-q)^2\right]\rho
             +Nc{v^2\over 6}(1-q)\rho^2 +q E^{\rm T} \nonumber\\
             &+&Nk_BT \ln \left[ \left({\pi\over A}\right)^{3/2}
                                     {q\over v}\right]
              +{\rho v\beta b^2 \over 4}2^{-3/2} q(1-q)^2
\end{eqnarray}
where $q \sim 1$. The first two terms, as is usual, have the form of a virial 
expansion, the third and fourth terms represent the enthalpy and entropy change 
due to folding, respectively. 
The last term, coming from inter-replica interaction, tends to cancel out 
the effective attraction 
due to the randomness appeared in the first term because protein does not 
feel randomness when it precisely coincides with the native structure. 

In summary, eqs.(\ref{eq:FCG}), (\ref{eq:Fglass}), and (\ref{eq:Ffolded}) 
are the expressions for the free energy for 
all relevant phases, which will be used in the next sections.

\section{PHASE TRANSITIONS AND FREE ENERGY LANDSCAPE}
\label{sec:phase}
Since we have obtained simple enough expressions for the free energies of several phases, we can
now discuss the `phase transitions' for finite systems. 
Our emphasis is on description of ruggedness of the free energy landscape. It has been 
argued repeatedly that there are a number of minima in the glass phase. 
We emphasize here, however, that even above the (static) glass transition 
temperature the appropriate free energy landscape has many minima which affect folding kinetics 
drastically.

\subsection{Coil-globule transition (collapse)}
We first discuss the coil-globule phase transition based on the free 
energy expression eq.(\ref{eq:FCG}) as a function of density $\rho$, $(\rho \geq 0)$. 
First of all, we ignore the third term, which is smaller in $N$. 
Then, the lowest free energy is attained at the $\rho^*=0$ when 
$b_{\rm eff}\equiv b_0-\beta b^2/2>0$, while it becomes positive, 
\begin{equation}\label{eq:rhostar}
\rho^*=-{3\over 2cv} \left( b_0-{\beta b^2\over 2}\right)
\end{equation}
when $b_{\rm eff}<0$. Thus, the phase transition temperature $T_{\rm CG}$ is determined by
\begin{equation}
b_0 - {\beta_{\rm CG}b^2\over 2} =0,
\end{equation}
where $\beta_{\rm CG}=1/(k_B T_{\rm CG})$. 
The third term in the free energy (\ref{eq:FCG}) does not change this temperature 
significantly for sufficiently large polymer. 
There are two cases, however, where the third term play roles.
First, for a short polymer at high temperature, the third term becomes 
dominant; this term makes the globule state unstable and so the 
random-coil phase always appears in the limit of high temperature.
Second, in the vicinity of $\rho=0$, the third term is the largest
and thus $\partial F /\partial \rho |_{\rho=0}$ is positive infinite and so 
the transition is first order with very small barrier $O(N^{-1})$. 
We should mention that extending the argument to include non-uniform description of polymer leads 
us to a surface term of order $O(N^{2/3})$\cite{Grosberg94}, which is not taken 
into account here. The third term here is $O(N^{1/3})$, which is smaller than
the surface term and so the reasoning leading to first order phase transition given here 
might not be appropriate. In any case, coil-globule transition is a first order phase 
transition with very small barrier and because of this it might be recognized as 
a second order like transition by numerical simulations, or in the laboratory.

\subsection{Globule-glass transition}
Next, we discuss the glass transition. First of all, we fix $\rho$ 
at $\rho^*$ given in eq.(\ref{eq:rhostar})\cite{Shakhnovich89}. 
Roughly, $\rho^*$ should not change significantly after the collapse although, 
rigorously speaking, $\rho$ should be optimized simultaneously with the other order parameters. 
Here, we should remember that our starting point was based on the virial expansion which is not 
very accurate in any collapsed phase. 
Thus we feel the virial approach will overemphasize density variations
$\rho^*$ in the virial approximation changes too rapidly with the other 
thermodynamic parameters that would be the case for a more accurate homopolymer 
equation of states.  
Therefore, it is better to fix $\rho^*$ by choosing $c$ appropriately at this 
level of description. In other words, we change from the independent parameter $c$ 
to $\rho^*$. Qualitative features do not change very much by this 
prescription. Again this will be a most accurate description when strong 
collapse is favored by the homopolymeric part of the pair interactions.

We seek the saddle solutions of $F_{\rm Glass}(m,X)$ (i.e., eq.(\ref{eq:Fglass}))
with respect to $m$ and $X\equiv (2mD)^{3/2}$.
First, let us minimize $F_{\rm Glass}/(m-1)$ with respect to $X$.
Forgetting the first two terms which are constant in $X$, we have 
two relevant terms which have opposite effects. The third term, the driving force 
to stabilize the replica symmetry breaking solution, tends to push $X$ to 
its maximum value $X_{\rm max}=(4\pi)^{3/2}v^{-1}$. 
See Fig.\ref{fg:FvsX}, in which a dotted line with `$T\ln X$' corresponds to the third term. 
On the other hand, the fourth term, the entropy loss due to freezing, prefers small $X$
(another dotted line with `$-\beta X$' in Fig.\ref{fg:FvsX}).
At sufficiently high temperature, the fourth term which is proportional to $T$ always 
dominates and so $X=0$ is the only stable solution, as is illustrated in the 
figure. As decreasing temperature, 
the third term which is inverse proportional to $T$ becomes important
at large $X$ and, in addition to the solution $X=0$, a new solution 
$X=X_{\rm max}$ becomes locally stable when
\begin{equation}
{\partial F_{\rm Glass}\over \partial X}\Big|_{X=X_{\rm max}}=0,
\end{equation} 
which gives
\begin{equation}\label{eq:defTA}
{\beta^2 b^2 \rho^* v m \over 4}=1.
\end{equation}
We call this critical temperature $T_{\rm A}$ following \cite{Kirkpatrick87b}.
For structural and Potts spin glasses this is the transition temperature 
predicted by mode coupling theory\cite{Kirkpatrick89}.
Below $T_A$, there are always two 
locally stable solutions $X=0$ and $X=X_{\rm max}$, 
a melted phase and frozen phase, respectively. 
We should mention that eq.(\ref{eq:Fglass}) is derived under the assumption that 
$D \gg A, B, C$, which does not hold true for the solution $X=0$ (i.e., $D=0$). 
Thus, we have to use the free energy expression for the coil-globule phase keeping 
a small dependence on $D$. We find $X=0$ is indeed a stable solution, at the end. 

Next, at $T \leq T_{\rm A}$, we optimize 
\begin{equation}\label{eq:Fvsm}
F(m,X_{\rm max})={1\over 4}N\rho^*v \left(b_0-{\beta b^2\over 2}\right)
                -{1\over 4}N\rho^* v\beta b^2(m-1)
                +Nk_BT{m-1\over m} (\ln p^\prime\gamma-3/2)
\end{equation}
with respect to $m$, where $p^\prime\equiv (8\pi)^{3/2}$ and $\gamma\equiv a^3/v$. 
($\gamma$ represents flexibility of the chain and is about 5 for very flexible chain 
like protein\cite{Grosberg94,flexibility}. The value of $p^\prime$ depends to some extent 
on the approximations we use and thus we think its precise value is 
somewhat uncertain. Qualitative 
results are not affected by its value as long as it is of order unity. 
In discussing lattice model results we therefore treat it as adjustable.)
In the same way as above, the second and third terms lead to effects in opposite directions
 (see Fig.\ref{fg:Fvsm}).
The stationarity condition,
\begin{equation}
{\partial F_{\rm Glass}(m,X_{\rm max})\over\partial m}=0,
\end{equation}
leads us to 
$m^*={2k_BT\over b} ({\ln p^\prime\gamma-3/2 / \rho^* v})^{1/2}$.
Inserting this into eq.(\ref{eq:defTA}) we get 
\begin{equation}\label{eq:TA}
k_BT_{\rm A}={b\over 2}\sqrt{\rho^*v(\ln p^{\prime}\gamma-3/2)}.
\end{equation}
At $T=T_{\rm A}$, $m_{\rm A}^*=\ln p^\prime\gamma -3/2$ is larger than unity and so 
in the ordinary replica formalism, this solution has been ignored 
for the reason that it does not contribute to the Boltzmann average. 
Physically this mean the configurational entropy of these local free energy minima 
is extensive at $T_{\rm A}$.
Recently, Kurchan, Parisi and Virasoro\cite{Kurchan93} interpreted this solution 
as yielding the metastable states in the case of $p$-spin spherical model. 
We follow their argument and allow $m$ to be larger than unity. 
The $m^*$ decreases in linear with $T$ and coincides with unity at $T=T_{\rm K}$ 
defined by
\begin{equation}\label{eq:TK}
k_BT_{\rm K}={b\over 2}\sqrt{\rho^* v\over \ln p^{\prime}\gamma-3/2}.
\end{equation}
Below this temperature, this frozen solution becomes dominant in the 
Boltzmann average. The Kauzmann temperature $T_{\rm K}$ corresponds to the case where 
the configurational entropy of basins reaches zero. 
Eq.(\ref{eq:TK}) is the same as that of Shakhnovich and Gutin\cite{Shakhnovich89} 
except that the estimates of the entropy loss, $k_B (\ln p^\prime\gamma-3/2)$ here, 
are not the same.
$T_{\rm K}$ is proportional to the randomness, $b$, 
and is inversely proportional to the square root of the entropy loss, 
which is the same dependence found by Bryngelson and Wolynes using 
a statistical field version of Flory theory\cite{Bryngelson87}.
Moreover, $T_{\rm K}$ is proportional to square root of $\rho^* v$ which represents packing 
fraction. This dependence is found in \cite{Plotkin96}. (At first glance, one 
may notice the difference by a factor $\sqrt{2}$ between the present result and that of 
Refs.\cite{Bryngelson87,Bryngelson95}. This is simply because of the difference in 
the definition of randomness as will be discussed later.)

Let us estimate the free energy at the saddle solutions. For the solution with $X=0$, 
\begin{equation}
F_{\rm Globule}^*=N{v\over 4}\left(b_0-{\beta b^2\over 2}\right)\rho^*,
\end{equation}
while for the solution with $X=X_{\rm max}$,
\begin{eqnarray}
F_{\rm Glass}^*&=&N{v\over 4}\left(b_0-{\beta b^2\over 2}\right)\rho^*\nonumber\\
               &-&Nb\sqrt{\rho^*v (\ln p^\prime\gamma-3/2)}
               +Nk_BT(\ln p^\prime\gamma-3/2)+{1\over 4}N\beta b^2 \rho^* v.
\end{eqnarray}
The energy difference between them has the simple form,
\begin{equation}
F_{\rm Glass}^*-F_{\rm Globule}^*=
(\ln p^\prime\gamma-3/2)Nk_BT_{\rm K}({T\over T_{\rm K}}+{T_{\rm K}\over T}-2)\geq 0,
\end{equation}
which touches zero at $T=T_{\rm K}$. It should be noted that in the replica 
formalism the solution with the larger free energy dominates the Boltzmann average when 
$m^*<1$ as is known\cite{Mezard87}, the solution with lower energy becomes dominant 
when $m^*>1$. Therefore, $F_{\rm Globule}^*$ dominates the thermal average above $T_{\rm K}$,
while $F_{\rm Glass}^*$ becomes dominant below $T_{\rm K}$.

Following Kurchan, Parisi, and Virasoro, we can estimate a lower bound for the 
free energy of 
transition states (TS) between the lowest local minima in the temperature range 
$T_{\rm A}>T>T_{\rm K}$. We conjecture that the behavior of the RSB in our 
model is analogous to that of $p$-spin spherical model and so the TS 
solution is again represented by 1 level RSB in the temperature range considering now.
Then, the TS solution $(m^\ddagger,X^\ddagger)$ may be assigned to the one which makes $F/(m-1)$ 
maximum with respect to $X$ ($X^\ddagger$ in Fig.\ref{fg:FvsX}). The saddle condition 
gives
\begin{equation}\label{eq:mddagger}
2 \ln {T\over T_{\rm A}} -\ln {m^\ddagger\over m_{\rm A}^*}+m_{\rm A}^*-m^\ddagger=0
\end{equation}
and
\begin{equation}\label{eq:Xddagger}
X^\ddagger={4(k_BT)^2(4\pi)^{3/2}\over m^\ddagger b^2 \rho^* v^2}.
\end{equation} 
The parameter $m^\ddagger >1$ indicates these are configurational entropy 
driven transitions. In general the parameter $m<1$ is conjugate to the 
nonextensive complexity of states below the thermal transition while here 
$m^\ddagger >1$ presumably represents the fact that multiple escape 
routes are possible from a trapped state.
The upper equation cannot be solved analytically in its general form. 
By the Taylor expansion around $T_{\rm A}$ ($T\leq T_{\rm A}$) we get
\begin{equation}
\Delta F^\ddagger\equiv F^\ddagger_{\rm Glass}-F_{\rm Glass}^*=Nk_BT_{\rm A}
  {m_{\rm A}^*-1\over m_{\rm A}^*+1}\left({T-T_{\rm A}\over T_{\rm A}}\right)^2
 +O\left[ \left({T-T_{\rm A}\over T_{\rm A}}\right)^3\right],
\end{equation}
which clearly shows that barrier heights grow up as decreasing temperature 
starting from zero at $T=T_{\rm A}$. 
Obviously, this temperature dependent barrier height will give a non-Arrhenius 
behavior in the kinetics, as is well-known in the structural glass physics.
Notice that this behavior is consistent with $T_{\rm A}$ being a sort of 
spinodal for the random minima.

We also show numerical estimate of the barrier heights at $T_{\rm A}>T>T_{\rm K}$ 
in Fig.\ref{fg:Fdagger}. 
We solve eq.(\ref{eq:mddagger}) numerically for $m^\ddagger$ and put it, 
together with eq.(\ref{eq:Xddagger}), into eq.(\ref{eq:Fglass}).
Here, barrier heights grow near $T_{\rm A}$ as was shown above 
and then start decreasing. The latter is because $X^\ddagger$ in eq.(\ref{eq:Xddagger}) 
decreases with temperature decreasing and the approximation $2mD\gg A, B, C$ becomes worse.
It is expected that real barrier heights grow monotonically. 
The value of barrier height depends on the non-universal number $p^\prime$
and may not be accurate;  
The naive choice of $p^\prime=(8\pi)^{3/2}$ gives $\Delta F^\ddagger\sim 1.0Nk_B T_{\rm K}$ at 
around $T_{\rm K}$, 
which is three times higher than that estimated from 27-mer lattice model 
by simulation\cite{Socci96}.
A smaller value $p^\prime=(8\pi)^{3/2}/5$ gives a comparable barrier height to the 
simulation $\sim 0.4Nk_BT_{\rm K}$. 
We should also note that the barrier height is always proportional to the size of 
polymer $N$ in the present description, which might be appropriate for relatively 
small proteins but is probably not accurate for larger ones where inhomogeneous 
saddle points may dominate. This may also be the reason the naive estimate gives
a larger barrier than the simulation. We will touch upon the 
latter case in the Discussions.

\subsection{Globule-folded transition (fast folding)}
To consider the folding transition we need a free energy expression 
applicable in the whole range $0\leq q \leq 1$. As discussed above, the entropic 
term in the globule phase is small and is negligible as the lowest approximation. 
Thus we simply 
interpolate the entropic term between two regimes, i.e., $q\sim 0$ and $q\sim 1$. 
Thus one uses a simple form:\cite{ignore},
\begin{eqnarray}
F_{\rm CGF}(q)&=&N{v\over 2}\left[b_0(1-q)-{\beta b^2\over 2}(1-q)^3\right]\rho
             +Nc{v^2\over 6}(1-q)\rho^2 +q E^{\rm T} \nonumber\\
             &+&Nk_BT \ln \left( p^{\prime\prime}\gamma q+1\right),
\end{eqnarray}
where $p^{\prime\prime}=(2\pi)^{3/2}$, the value of which should not be taken as very precise.
We again fix $\rho$ at $\rho^*$ given by eq.(\ref{eq:rhostar}) by choosing $c$ 
appropriately. 
Typical free energy curves along order parameter $q$ are drawn in Fig.\ref{fg:Fvsq}, 
which clearly shows that the folding transition is the first order. 
Thus, the globule-folded phase transition is defined by the relation,
\begin{equation}
F_{\rm CGF}(0)=F_{\rm CGF}(1),
\end{equation} 
which now gives
\begin{equation}\label{eq:TF}
-{1\over 8}\rho^*v \beta_{\rm F}b^2 =
-|\delta\epsilon^{\rm T}|+k_BT_{\rm F} \ln p^{\prime\prime}\gamma,
\end{equation}
where  $T_{\rm F}$ is the folding temperature,
 $\beta_{\rm F}=1/(k_BT_{\rm F})$.
We also defined the energy gap (per monomer) $\delta\epsilon^{\rm T}$ 
between the native energy and the average energy of collapsed states by 
\begin{equation}\label{eq:gap}
\delta\epsilon^{\rm T}=E^{\rm T}/N-\rho^*vb_0/4, 
\end{equation}
since the energy gap $\delta\epsilon^{\rm T}$ is more useful parameter than $E^{\rm T}$ 
to represent the bias towards folding\cite{on_gap}.
The LHS of eq.(\ref{eq:TF}) is the free energy of the globule, the first term in the RHS is the 
energy in the native state, and the second term is the entropy loss due to the 
folding. 

The critical situation at which free energy barrier for the folding transition
 disappears is determined by the relation 
\begin{equation}
{\partial F_{\rm CGF}\over \partial q}\Big|_{q=0}=0,
\end{equation}
which now leads to
$|\delta\epsilon^{\rm T}|=k_BT_{\rm D} p^{\prime\prime}\gamma
                  +{5\over 8}\rho^*v \beta_{\rm D}b^2.$
Here $T_{\rm D}$ is the critical temperature below which downhill 
folding occurs and $\beta_{\rm D}=1/(k_B T_{\rm D})$.

Above this critical temperature $T_D$, folding takes place over a free 
energy barrier. To think about it, we rewrite $q$-dependent part of free energy as 
$$
F_{\rm CGF}/N=\left(\delta\epsilon^{\rm T}-{1\over 8}\rho^*v\beta b^2\right)q 
             -{1\over 4}\rho^*v\beta b^2 (1-q)^3 +k_BT\ln (p^{\prime\prime}\gamma q+1)
             +{\rm const}.
$$
The first term, mainly representing the enthalpy change due to the folding, 
is linearly decreasing with respect to $q$. On the other hand, both the 
second term, the effective attraction due to the randomness, and the third term 
representing the entropy loss through the folding are monotonically 
increasing function of $q$. Therefore, the physical origin of the barrier 
for the folding is partly the reduction of randomness upon folding and partly the entropy loss; 
Depending on the values of parameters, either one can be dominant.
Because of this complication, the estimate of barrier height becomes 
complicated too. When the barrier is small, we can write it as
$$
\Delta F^\ddagger \sim N {\left(\delta \epsilon^{\rm T}+5\rho^*v\beta b^2/8
                          +K_BTp^{\prime\prime}\gamma\right)^2\over 
                          3\rho v \beta b^2+2k_BT (p^{\prime\prime}\gamma)^2}.
$$
Note that in this analysis the barrier height is always proportional to $N$, 
as is in the case of the barrier between lowest misfolded states discussed before.

\subsection{Glass-folded transition}
\par\noindent
First, we fix $\rho$ at $\rho^*$ by eliminating $c$ through eq.(\ref{eq:rhostar}), as usual.
The first order phase transition between the glass phase and the folded phase is defined 
by 
\begin{equation}
F^*_{\rm Glass}=F_{\rm CGF}(1),
\end{equation}
which now becomes
\begin{equation}
-|\delta\epsilon^{\rm T}|+k_B T_{\rm F}^\prime \ln p^{\prime\prime}\gamma
={1\over 8}\rho^* v \beta_{\rm F}^\prime b^2
-b\sqrt{\rho^* v (\ln p^\prime \gamma -3/2)} 
+k_B T_{\rm F}^\prime (\ln p^\prime\gamma-3/2),
\end{equation}
where $T_{\rm F}^\prime$ is the folding temperature from the glass phase
 and $\beta_{\rm F}^\prime=1/(k_B T_{\rm F}^\prime)$.

\section{DISCUSSIONS}
\label{sec:disc}
\par\noindent

\subsection{Phase diagram}

Taking results in the previous section together we can draw a few 
phase diagrams which are given in Figs.\ref{fg:diagram}. The present model includes 
4 independent parameters of interest, the mean value of contact energy $b_0$, 
variance of a contact interaction $b$, the energy gap between the native 
energy and the mean energy of collapsed states $|\delta\epsilon^{\rm T}|$ (defined in 
eq.(\ref{eq:gap}) ), 
and temperature $T$ and so we have no choice but to draw 
a few surface of sections of the complete four dimensional phase diagram.
As was mentioned before, we are not taking some numerical factors $p$, $p^\prime$, 
and $p^{\prime\prime}$ very literally at this level of description and we use 
the values deduced from lattice model, as will be explained below. 
A more sophisticated treatment is required to decide these values without the 
use of simulation data.
We also note that the qualitative features of the phase diagram do not change with 
the choice of these parameters so long as they are of order unity.

Figure \ref{fg:diagram}a shows a surface of section on $b-|\delta\epsilon^{\rm T}|$ 
plane, which is the same representation as that of Bryngelson and Wolynes\cite{Bryngelson87}.
In the figure, there are three phases (bounded by solid curves), the globule phase denoted by M, 
the glass phase denoted by G and the folded phase denoted by F. Roughly
speaking, the phase diagram at this level is essentially the same as that of 
Bryngelson and Wolynes. Now, we can go forward to put in more information on 
the kinetics. First, the globule phase is separated into two regimes, $\rm M_1$ regime 
where the free energy landscape is monotonous and no local minimum except globule 
state exists and $\rm M_2$ regime where there are many local minima, each of which 
is separated by a barrier of order $N$ though the formal Boltzmann average 
is dominated by the globule state. 
Next, the folded phase can be separated into three parts. These are the $\rm F_1$ regime where 
protein must pass over a free energy barrier to fold but is not trapped by the 
frozen state, the $\rm F_2$ regime where the protein does not experience an activation barrier 
for the folding transition and fast downhill folding occurs, and the 
$\rm F_3$ regime where before reaching at the folded state, the protein 
can be found in the glassy state and thus corresponds to a slow folding process 
with intermediate.
In the recent synthesis by Bryngelson et al\cite{Bryngelson95}, several 
scenarios of folding were classified. The Type 0 scenario there 
corresponds to the $F_2$ regime here, the Type I scenario takes place in
the $F_1$ regime, and the Type II scenario roughly corresponds to the 
$F_3$ regime. 
For the latter case, Bryngelson et al discussed the case where glass transition 
takes place at the middle of folding process assuming that glass transition temperature 
$T_{\rm K}(q)$ increases as a function of $q$. 
This seems to be the case in the lattice models studied\cite{Onuchic95}.
The present analysis, however, suggests the possibility of 
the opposite situation, i.e., the glass transition temperature decreases as a function of 
$q$ and thus there can be a case where the pre-folding state is glassy, while 
folded state is non-glassy. There is indeed a subtle issue whether the glass 
transition temperature increases or decreases as a function of $q$ because 
both the ruggedness $\Delta E^2$ and the residual entropy $S$ decrease with respect to $q$.
The latter quantity depends on the variational approximation used.

Figure \ref{fg:diagram}b shows another surface of section on the $k_BT-b$ 
plane (when $b_0$ is negative). This representation  
corresponds to that of Sasai and Wolynes\cite{Sasai90}. 
The same notation as above is used to describe each phase/regime. 
It is well-known that the temperature dependence of folding is not simple to discuss in 
laboratory studies of proteins because every parameter, in principle, depends 
on temperature through the entropic contribution to the hydrophobic force. 
A similar problem occurs when comparing the phase diagram for the virial expansion 
Hamiltonian to a lattice model with rigid excluded volume.
In order not to make the argument ambiguous we first ignore 
any dependence of the parameters $b_0$, $b$, and $|\delta\epsilon^{\rm T}|$ 
on temperature. The phase diagram looks similar to that found by Sasai and Wolynes. 
One outstanding difference is that, in the present diagram, we have no random-coil phase,
 which appeared in Sasai and Wolynes. One reason is
that we have ignored the entropic term to locate the coil-globule transition.  
As was mentioned, the entropic term always creates the coil phase in the 
high temperature limit.
The other reason is related to the difference in the model itself; 
We assumed that all parameters are independent of $T$, for clarity and 
this is why we have no coil phase in this surface of section. 

To be more realistic, we next consider the temperature dependence of the average 
virial coefficient $b_0$, which 
leads us to the random-coil phase in this representation.
To see this, we employ the temperature dependence of 
$b_0$ as
\begin{equation}\label{eq:b0change}
b_0={T-\theta\over \theta}2k_B T
\end{equation}
following Grosberg and Khokhlov\cite{Grosberg94}. Here, $\theta$ is the so-called 
theta-temperature where $b_0$ becomes zero. The phase diagrams for this model 
are given in Figs.\ref{fg:diagram2} for two different values of $\theta$. 
This has the random-coil phase denoted by C as
is expected and is closer to that of Sasai and Wolynes\cite{Sasai90} 
as well as that of Bryngelson et al\cite{Bryngelson95}.
For a better solvent (Fig.\ref{fg:diagram2}b), the coil phase becomes more stable and thus 
the coil-globule transition curve goes down. Depending on the nature 
of solvent, a direct transition from the coil phase to the folded phase 
may also occur. 

We now comment upon relations to other phase diagrams given in literature. 
Shakhnovich and Gutin\cite{Shakhnovich89} showed a phase diagram on $b-b_0$ plane 
for the random heteropolymer.  Restricting $C\equiv 0$, we can compare 
two results quantitatively. 1)The coil-globule 
transition curve is exactly the same. 2)Comparing eq.(\ref{eq:TK}) with eq.(26) of 
\cite{Shakhnovich89} we see that the only difference arises numerical factors which are 
not very exact in either analysis.
Next, we comment on the phase diagram given in Ramanathan and Shakhnovich
\cite{Ramanathan94}. Roughly speaking, the selective temperature there plays 
similar role to $b/|\delta\epsilon^{\rm T}|$ here. Thus, interchanging the 
vertical and horizontal axes, we see that Fig.\ref{fg:diagram}b and Fig.1 of 
\cite{Ramanathan94} look very similar. 
Socci and Onuchic drew a phase diagram based on the lattice MC simulation. 
Unfortunately, we cannot compare directly with their results because they 
fix the sequence while changing interactions between monomers, thus both $b_0$ and 
$|\delta\epsilon^{\rm T}|$ depend on interactions simultaneously.

\subsection{Free energy landscape}

We can get some insight on the ruggedness of free energy landscape based on the analogy between 
the Potts type spin glass and the present model. 
Figure \ref{fg:TAP} represents schematically the TAP free energy in the Potts-type 
spin system. This can also be viewed as a TAP free energy landscape of the random heteropolymer,
although we do not present here any explicit form of TAP free energy; 
We can define `pure states' if the potential surface has many minima, each of which is 
separated from the others by an infinite barrier in the thermodynamic limit. 
An individual pure state $s$ can be identified with the expectation value $\{\bar{\bf r}_i^s \}$ 
of monomers averaged for a particular local minimum. 

Referring Fig.\ref{fg:TAP} we discuss characteristics of the free energy landscape for 
each temperature range.
1)At any temperature $T$ above $T_{\rm A}$, the landscape is monotonous and 
there is only one trivial solution in TAP equation. 
In the replica formalism, this is represented by the replica-symmetric solution 
(the solution with $D=0$),
which corresponds to the globule state physically. 
2)At the temperature between 
$T_{\rm A}$ and $T_{\rm K}$, there are both replica-symmetric and RSB solutions.
The latter energy coincides with the lowest TAP free energy. To account for the 
formal Boltzmann average we sum over many TAP solutions as
\begin{equation}\label{eq:TAPsum}
Z=\int dF_{\rm TAP} \exp \left[ \ln \omega(F_{\rm TAP})-\beta F_{\rm TAP} \right]
\end{equation}
where $\omega$ is the density of TAP solutions. In this temperature range, the 
exponent here has the stationary point $F^*$ above the lowest TAP energy. 
Therefore, a number of TAP solutions contribute to the Boltzmann average and, 
due to this degeneracy (complexity), the free energy $F=-1/\beta \ln Z$, which 
coincides with the replica-symmetric free energy $F_{\rm RS}$, becomes lower 
than the lowest TAP energy ($=F_{\rm RSB}$). 
For finite systems such as actual proteins, the infinitely long time behavior can 
be represented by the replica-symmetric solution, i.e., globule state, while 
the protein nevertheless feels a large barrier to move between globally different 
states and finite time dynamics may be 
controlled by the metastable RSB solution, i.e., glassy state. Activated 
transport among many TAP minima takes place. 
3)Below $T_{\rm K}$, the stationary point $F^*$ in the exponent of 
eq.(\ref{eq:TAPsum}) disappears and a few lowest minima in the free energy landscape 
dominate the Boltzmann average. Since the lowest TAP free energy still coincides with 
the RSB free energy, the glassy state becomes globally stable.

Next, we discuss the physical interpretations of $T_{\rm A}$ and $T_{\rm K}$. 
$T_{\rm K}$ is the temperature where the entropy becomes zero and sometimes 
called `entropy crisis'. 
The average number of contacts can be estimated as $\rho^* v/2$ and 
variance of random energy distribution per monomer 
becomes $\Delta \epsilon^2\sim \rho^*vb^2/2$, while 
the entropy loss per monomer $s_{\rm loss}$ through freezing is $k_B(\ln p^\prime\gamma-3/2)$. 
Therefore, $T_{\rm K}$ is expressed as 
$$
k_BT_{\rm K}=\sqrt{(\Delta \epsilon)^2\over 2 s_{\rm loss}/k_B},
$$
which is consistent with the analysis of the REM\cite{Bryngelson95,Bryngelson87}.
Note that the entropy factor is in the denominator; 
Increasing {\em the flexibility per Kuhn segment} $\gamma$ \cite{flexibility}, 
the entropy to be lost for freezing 
increases and then the Kauzmann temperature $T_{\rm K}$ becomes lower. 
On the other hand, since $T_{\rm A}$ is given by 
$k_BT_A={b\over 2} \sqrt{\rho^* v (\ln p^\prime\gamma-3/2)}$,
increasing the flexibility $\gamma$, it becomes easier to make multiple minima in the 
free energy landscape and thus $T_{\rm A}$ increases. 
As a result, a polymer with large $\gamma$ possesses a relatively wide temperature 
range between $T_{\rm A}$ and $T_{\rm K}$.

In the previous section, we have shown that barrier heights between two 
lowest minima in the free energy landscape increase with decreasing temperature 
below $T_{\rm A}$. This directly leads to the super-Arrhenius temperature 
behavior, for example, in the diffusion constant in this temperature regime. 
On the other hand, from the analogy to the REM\cite{Bryngelson89}, we expect
that barrier heights saturate at $T_{\rm K}$, below which the diffusion constant 
recovers Arrhenius temperature dependence. 
To deal with the temperature range below $T_{\rm K}$ explicitly, 
following Ref.\cite{Kurchan93}, we need to employ 2 level RSB, which is straightforward 
but somewhat more elaborate.

\subsection{Mapping onto 27-mer lattice model of protein}

We can try to map the present model onto the three-letter code 27-mer 
lattice model studied exhaustively 
by MC simulations\cite{Onuchic95,Socci96} 
although the present model is not a lattice-type one. 
While this mapping must be ambiguous to some extent, it is helpful 
in understanding the simulation results. 
In the lattice model\cite{Onuchic95,Socci96}, 
1)the average energy of globule state is $-50$, which corresponds 
to $\rho v b_0/4\leftrightarrow -50/27=-1.8$. 
2)The energy of the native structure is $-84$ and thus 
subtracting the average energy, we get the stability gap per monomer, 
$|\delta \epsilon^{\rm T} | \leftrightarrow (84-50)/27= 1.3$. 
3)The entropy loss for freezing to an unique structure from a free chain 
should be roughly $\ln 5$ for the cubic lattice giving the estimate 
$k_B(\ln p^\prime \gamma -3/2)\leftrightarrow 1.6 $. 
4)On the other hand, the entropy loss for globule-folded 
transition was estimated as $20k_B$ from simulation 
of 27-mer at the folding temperature giving a different value
$\ln p^{\prime\prime}\gamma\leftrightarrow 20/27 = 0.74$.
5)The measured energy fluctuation at the folding temperature is 51 and so 
$\rho v b^2/2 \leftrightarrow 51/27= 1.88$.
From these mappings, we can get the dynamic glass temperature, the static glass 
temperature, and the folding temperature to be $T_{\rm A}=1.23$, $T_{\rm K}=0.77$, 
and $T_{\rm F}=1.25$, respectively. On the other hand, there is no real solution 
for $T_{\rm D}$; Apparently a strictly downhill scenario folding 
(spinodal folding) cannot be reached with these parameters. 

These estimates are very crude and tentative, but it still may be interesting to 
discuss the folding scenario based on it. From this estimate $T_{\rm A}$ and $T_{\rm F}$ 
are very close to each other and thus, usually folding occurs below $T_{\rm F}$, 
so the protein feels multiple minima along its folding route. 
Therefore, the most probable scenario is that, after hopping among many local minima, 
protein finds its native structure which is stable thermodynamically. 
On the other hand, because the estimate are uncertain, one should consider the 
possibility that folding may occur in a regime where chain dynamics is not 
far from Rouse dynamics renormalized by mode coupling effects\cite{Schweizer89}.

We should give a warning that changes in the mapping procedure may change 
this assignment of the scenario to some extent. 
Folding is expected to occur at temperatures somewhat above $T_{\rm K}$ and thus 
the characteristics of free energy landscape in this regime are of most interest.
More exhaustive study of off-lattice simulation 
seems to be desirable to study details of free energy landscape and kinetics.
The present analytical calculation should be tested more carefully by off-lattice simulation.

Very recently, Shakhnovich and co-workers have carried out lattice simulations that 
suggest random heteropolymer dynamics is not activated at high 
temperature\cite{Shakhnovich96}, adducing 
a polynomial dependence of the time scale on system size. 
This would be consistent with the expectation of a mode coupling analysis about $T_{\rm A}$, 
where polynomial divergences with chain length are expected\cite{Gotze91}. 
At low temperatures activated dependence is seen, however. A more exhaustive version 
of such studies may help in quantifying $T_{\rm A}$ versus $T_{\rm K}$.

\subsection{Folding kinetics: Comments on the effects of inhomogeneity}

In this paper we concentrated on the case where all order parameters $B$, $C$, 
and $D$ are independent of $i$; We have restricted our description to 
a homogeneously ordered or trapped polymer. 
In particular, because of this the two types of barrier heights discussed 
in this paper are proportional to $N$, which may not be appropriate for 
relatively large protein. For the latter, many inhomogeneous states may 
play important roles, especially to describe the folding kinetics. 
First, a folding nucleus can be represented as a state where part of 
the chain has much larger $C_i$ than the other part. 
Nucleation\cite{Bryngelson90,Abkevich94,Guo95,Thirumalai95} 
would naturally be followed by the growth of the number of monomers having larger $C_i$.
The question addressed by such a analysis would be the size 
dependence of the free energy barrier for the folding transition.
Secondly, for quite a large polymer, a specific separation of $C_i$ into two values may 
create a (meta)stable state. This might be related to a foldon, a small 
quasi-independent folding unit\cite{Panchenko95}. 
Another possibility of inhomogeneous states is a locally trapped state, 
which means only part of chain has large $D_i$, while the others have $D_i\sim 0$.
This might be a transition state between totally frozen states and that 
between a frozen state and melted state.
We can extend our treatment to these inhomogeneous cases with the trial Hamiltonian,
\begin{eqnarray}
\beta H_{\rm ref}&=&A \sum_{\alpha ,i}({\bf r}_{i+1}^\alpha-{\bf r}_{i}^\alpha)^2
  +\sum_{\alpha ,i} B_i ({\bf r}_i^\alpha)^2
  +\sum_{\alpha ,i} C_i ({\bf r}_{i}^\alpha-{\bf r}_{i}^{\rm T})^2
  +\sum_{\alpha\neq\beta ,i}D_i d_{\alpha\beta}({\bf r}_{i}^\alpha-{\bf r}_{i}^{\beta})^2.
\end{eqnarray}
Detailed results for these problems are under study and will be 
reported elsewhere.

These inhomogeneous description of polymer require more 
 elaborated modes of replica symmetry breaking. For example, in the case of 
$p$-spin spherical model, 1 level RSB is enough to represent stable states 
at temperature below $T_{\rm K}$, while at the same temperature range 2 level 
RSB describe transition states between two lowest minima\cite{Kurchan93}.
Correlation of the free energy landscape has been modeled in terms of generalized 
REM, which includes continuous part of Parisi's order parameter and is a step 
to this direction\cite{Plotkin96}.

We note here that an inhomogeneous version of the analysis here gives 
results analogous to Kirkpatrick 
and Wolynes's first calculation of barriers based on a simple interface 
between TAP solutions\cite{Kirkpatrick87b}. 
Indeed Parisi\cite{Parisi94} obtained the same dependence as KW on 
$(T-T_{\rm K})$ i.e., $\Delta F^\ddagger\sim (T-T_{\rm K})^{-2}$ for 
Potts glasses in 
3 dimensions. Later, Kirkpatrick, Thirumalai, and Wolynes showed how 
wetting of the interface between two TAP solutions in a droplet 
by other TAP solutions led to the more 
usual Vogel-Fulcher behavior $\Delta F^\ddagger\sim (T-T_{\rm K})^{-1}$
\cite{Kirkpatrick89}b).
This would apparently require a more complete spatially inhomogeneous 
RSB for the saddle point. We note that recently Thirumalai has argued, based 
on the KTW style argument, that barriers for traps should scale only as 
$N^{1/2}$\cite{Thirumalai95}. 
In our view further analysis is needed because this scaling argument 
should be only valid in the strict vicinity of $T_{\rm K}$, not necessarily 
the high temperature relevant for folding of minimally frustrated proteins.

As was noted, virial expansion used in this paper is not very appropriate 
to describe compact states. Incorporating rigid chain connectivity with hard core repulsion 
 will overcome this disadvantage. In principle, Fixman's independent-
oscillator comparison potential\cite{Fixman69} might be applied to do this, although 
qualitative results discussed here are believed to be unchanged. 
The effects of the hardcore on $T_{\rm A}$ may be considerable, since even the 
homogeneous hard sphere fluid possesses a mean field dynamical transition. 
Also the somewhat subtle questions such as change in the radius of gyration from 
molten-globule state to the folded state may be analyzed by this extension.

\section{CONCLUSIONS}
\label{sec:conc}
We have analyzed the free energy landscape of model protein based on the 
replica variational method. The ruggedness of free energy landscape is manifested
by two glass transition temperatures, $T_{\rm A}$ and $T_{\rm K}$ ($T_{\rm A}>T_{\rm K}$). 
1)Above $T_{\rm A}$, the landscape is monotonous. Dynamics is like that 
of a free Rouse chain modified by mode coupling effects. 
These effects would give dynamical freezing at $T_{\rm A}$. 2)Between $T_{\rm A}$ and 
$T_{\rm K}$, the landscape has a number of metastable minima but the collection 
of them dominates the Boltzmann average as a whole. These are represented by the 
replica symmetric solution, while the RSB solution is
metastable. 3)Below $T_{\rm K}$,  only a few lowest states contribute to 
the Boltzmann average and this is well-described by the RSB solution. 
In the second regime, the barrier between two lowest minima grows as decreasing 
temperature, which leads to the super-Arrhenius temperature dependence of diffusion 
constant. 
We believe that folding occurs somewhat above $T_{\rm K}$, which implies a 
mechanism of hopping between numerous local minima until finding native structure 
through the guiding forces provided by minimal frustration and the concomitant 
folding funnel. 
We have also drawn a phase diagram having seven qualitatively different dynamical 
regimes. Several scenarios of folding were discussed based on this diagram 
although it may not be quantitatively accurate in all details.

\section*{ACKNOWLEDGMENTS}
We thank Masaki Sasai and Eugene I. Shakhnovich for discussions of their results. 
S.T. is supported by the Postdoctoral Fellowships for Research Abroad of Japan 
Society for the Promotion of Science. 
P.G.W.'s work on folding is supported by NIH grant PHS 1 R01 GM44557.
\par\noindent

\appendix
\section{EXPLICIT EXPRESSIONS}
\par\noindent
Here we give explicit expressions for the variational free energy $F_{\rm var}$ 
defined in eq.(\ref{eq:var-ineq}) in terms of parameters $B$, $C$, $D$ and $m$ 
that appear in the reference Hamiltonian.

Since $Z_{\rm ref}$ is a many dimensional Gaussian integral with exponent 
$A\sum_{\mu ij}{\bf r}^\mu_i{\cal H}^\mu_{ij} {\bf r}^\mu_j$, we can execute the integration 
by calculating the inverse matrix of ${\cal H}^{\pm}$,
\begin{eqnarray}
G_{ij}^\pm&=&\left\{ \begin{array}{ll}
         {\displaystyle{\cosh (i-1/2)\lambda_\pm \cosh (N-j+1/2)\lambda_\pm} \over 
          \displaystyle{\sinh N\lambda_\pm \sinh \lambda_\pm}}  & i < j \\
         {\displaystyle{\cosh (j-1/2)\lambda_\pm \cosh (N-i+1/2)\lambda_\pm} \over 
          \displaystyle{\sinh N\lambda_\pm \sinh \lambda_\pm}}  & i \geq j
            \end{array}\right.
\end{eqnarray}
and the determinant, where $\lambda_{\pm}$ is defined by
\begin{equation}
\sinh \lambda_\pm ={f_\pm \over 2A}\sqrt{1+4A/f_\pm},\qquad 
\cosh \lambda_\pm =1+{f_\pm\over 2A},
\end{equation}
where $f_\pm=B+C+\Lambda_\pm D$ and $\Lambda_+$=0, $\Lambda_-=2m$.
Using these, we can write down $Z_{\rm ref}$ as
\begin{eqnarray}
Z_{\rm ref}&=&\left( \pi\over A \right)^{3n(N-1)/2}N^{-3n/2}
   \left( \sinh \lambda_+\over \sinh N\lambda_+ \right)^{3n/2m}
   \left( \sinh \lambda_-\over \sinh N\lambda_- \right)^{3n(m-1)/2m}\nonumber\\
  &\times&  \exp \left( {nC^2\over A}\sum_{ij}{\bf r}_i^{\rm T} G_{ij}^+ 
   {\bf r}_j^{\rm T} - nC\sum_i {\bf r}_i^{\rm T 2}\right).
\end{eqnarray}

As was explained in Sec.\ref{sec:replica}, the conformational entropy $S$ 
expressed as
\begin{eqnarray}
S/k_B&=& -\beta F_{\rm ref} + \beta \langle H_{\rm ref} \rangle 
      -A\sum_{\alpha,i}\langle ({\bf r}_{i+1}^\alpha-{\bf r}_{i}^\alpha)^2
       \rangle,
\end{eqnarray} 
can be computed directly from $Z_{\rm ref}$ and is written as
\begin{eqnarray}
     &=& {3n\over 2m}\left( 1+A{\partial\over\partial A}\right)
         \ln {\sinh \lambda_+\over \sinh N\lambda_+}
     +{3n(m-1)\over 2m}\left( 1+A{\partial\over\partial A}\right)
         \ln {\sinh \lambda_-\over \sinh N\lambda_-}\nonumber\\
     && -n{C^2\over A} \sum_{ij}{\bf r}_i^{\rm T}
    \left(1-A{\partial\over\partial A}\right)G_{ij}^+{\bf r}_j^{\rm T},
\end{eqnarray} 
where we dropped a trivial constant term which represents Gaussian free chain entropy.

For the one-replica part, $\langle H_1 \rangle$ is written in terms of 
$\langle \rho_\alpha\rangle$ and $\langle q \rangle$ in the text. Thus, what 
we need to do here is to give expressions for the latter two quantities. 
The expressions for these just 
have an additional $\delta$-function from the definition of $Z_{\rm ref}$ and 
are easily computed to give
\begin{equation}
\langle \rho ({\bf r})\rangle =\sum_i \left({\pi g_i\over A}\right)^{-3/2}
\exp \left[ -A({\bf r}-{\bf s}_i)^2/g_i\right],
\end{equation}
where 
\begin{equation}
g_i \equiv {1\over m}\left[ G_{ii}^+ +(m-1)G_{ii}^- \right],
\end{equation}
\begin{equation}
{\bf s}_i \equiv {C\over A}\sum_l G_{li}^+ {\bf r}_l^{\rm T}, 
\end{equation}
and
\begin{equation}
\langle q \rangle ={v\over N}\sum_i \left({\pi g_i\over A}\right)^{-3/2}
\exp \left[ -A({\bf r}_i^{\rm T}-{\bf s}_i)^2/g_i\right].
\end{equation}

Finally, the inter-replica term $\langle H_2\rangle$ is written in terms of 
$\rho_\alpha$ and $\langle Q_{\alpha\beta} \rangle$. The latter is computed as
\begin{eqnarray}
\langle Q_{\alpha\beta}\rangle
           &=& \left\{ \begin{array}{l} 
\displaystyle{\sum_i \left({\pi \tilde{g}_i\over A}\right)^{-3/2}
                          \left({\pi G_{ii}^-\over A}\right)^{-3/2}
        \exp \left[-{2A\over \tilde{g}_i}\left({{\bf r}_1+{\bf r}_2\over 2}
                                               -{\bf s}_i\right)^2 
                   -{A\over 2G_{ii}^-}({\bf r}_1-{\bf r}_2)^2\right]} \\
\displaystyle{\sum_i \left({\pi g_i\over A}\right)^{-3} 
        \exp \left[-{A\over g_i}({\bf r}_1-{\bf s}_i)^2 
                   -{A\over g_i}({\bf r}_2-{\bf s}_i)^2\right]}, 
               \end{array} \right.
\end{eqnarray}
where upper (lower) line is for the case $\alpha$ and $\beta$ belong to 
the same (different) group of RSB and $\tilde{g}_i$ is defined by
\begin{equation}
\tilde{g}_i ={2\over m} G_{ii}^+ +\left( 1-{2\over m}\right)G_{ii}^-.
\end{equation}

In the limit of large $N$, expressions become considerably simple. For $G_{ij}^{\pm}$
\begin{equation}
G_{ij}^\pm\simeq {1\over 2\sinh \lambda_\pm}\exp (-|i-j|\lambda_\pm).
\end{equation}
which depends only on the sequential distance $|i-j|$ between monomers. 
Therefore, $G_{ii}^\pm$ is independent of $i$ and so
\begin{eqnarray}
\langle Q_{\alpha\beta}\rangle
           &=& 
\displaystyle{\left({2\pi G_{ii}^-\over A}\right)^{-3/2} 
         \exp \left[-{A\over 2G_{ii}^-}({\bf r}_1-{\bf r}_2)^2 \right]
         \sum_i \left({\pi \tilde{g}_i\over 2A}\right)^{-3/2}
         \exp \left[-{2A\over \tilde{g}_i}\left({{\bf r}_1+{\bf r}_2\over 2}
                                               -{\bf s}_i\right)^2 \right]} \\
           &\sim& \rho_\alpha\left({{\bf r}_1+{\bf r}_2\over 2}\right) 
     \left({2\pi G_{ii}^-\over A} \right)^{-3/2}
     \exp \left[-{A\over 2G_{ii}^-}({\bf r}_1-{\bf r}_2)^2\right],
\end{eqnarray}
for the case $\alpha$ and $\beta$ belonging to the same group of 1 level RSB, and
\begin{eqnarray}
\langle Q_{\alpha\beta}\rangle
           &=& 
     \rho_\alpha\left({{\bf r}_1+{\bf r}_2\over 2}\right) 
     \left({\pi g_i\over A} \right)^{-3/2}
     \exp \left[-{A\over g_i}({\bf r}_1-{\bf r}_2)^2\right],
\end{eqnarray}
otherwise.


\par\noindent
\vfill
\begin{figure}
\caption{Schematic view of the TAP free energy landscape with the Boltzmann 
distribution plots. a)Above $T_{\rm A}$, 
the free energy landscape is monotonous. b)At $T_{\rm A}>T>T_{\rm K}$, 
the free energy landscape has a 
number of minima and a collection of metastable states contribute to the Boltzmann average, 
which corresponds to the replica symmetric solution $F_{\rm RS}$.
c)Below $T_{\rm K}$, the free energy landscape has a number of minima but only a few lowest 
states dominate the Boltzmann average, which is calculated 
by the RSB solution $F_{\rm RSB}$.}
\label{fg:TAP}
\end{figure}

\begin{figure}
\caption{The free energy as a function of $X=(2mD)^{3/2}$. The dotted curve 
with `$T\ln X$' represents 
the third term of eq.(\protect\ref{eq:Fglass}) and the dotted curve with 
 `$-\beta X$' corresponds to 
the fourth term. Three solid curves represent sum of them for different temperatures.
$T=T_{\rm A}$ is the critical temperature below which there is a minimum at $X=X_{\rm max}$.}
\label{fg:FvsX}
\end{figure}

\begin{figure}
\caption{The free energy as a function of $m$. The dotted curve with `$T (m-1)/m$' represents 
the second term of eq.(\protect\ref{eq:Fvsm}) and the dotted curve with 
`$-\beta (m-1)$' corresponds to 
the third term. Three solid curves represent sum of them for different temperatures.
$T=T_{\rm K}$ correspond to the critical temperature at which $m^*=1$.}
\label{fg:Fvsm}
\end{figure}

\begin{figure}
\caption{The free energy barrier between two lowest minima as a function of temperature. }
\label{fg:Fdagger}
\end{figure}

\begin{figure}
\caption{The free energy as a function of nativeness $q$ for 4 different temperatures. 
1)$T>T_{\rm F}$, 2)$T=T_{\rm F}$, the folding transition temperature, 3)
$T_{\rm F}>T>T_{\rm D}$ where folding occurs through 
activation process, and 4)$T=T_{\rm D}$ below which downhill folding takes place.}
\label{fg:Fvsq}
\end{figure}

\begin{figure}
\caption{Phase diagrams derived from the present model. a)a diagram on 
$b-|\delta\epsilon^{\rm T}|$ plane with energy unit $k_B T$. 
b)a diagram on $k_BT-b$ plane with energy unit $|\delta \epsilon^{\rm T} |$.
Solid curves separate different (static) phases, dotted curves represent 
the boundaries at which metastable states disappear, and dashed curve 
shows the glass transition in the metastable unfolded state.   
$\rm M_1$ region is the molten globule phase with monotonous free energy landscape. 
$\rm M_2$ region still corresponds to the molten globule state, but rugged 
free energy landscape has many minima. $\rm G$ means the glassy phase where protein misfold 
to any one of the lowest states. ${\rm F}_n$ with $n=1,2,$ and 3 are folded phase; 
$\rm F_1$ region has an energy barrier to folding but the system is not frozen, while there
is no barrier in $\rm F_2$ region. $\rm F_3$ corresponds to the regime where pre-folding state 
is glassy and thus folding transition can be very slow. The values of parameters used are 
$\rho v=1$, $\ln p^\prime \gamma-3/2=1.6$, and $\ln p^{\prime\prime}\gamma=0.74$, which are 
deduced from mapping onto 27-mer lattice simulation.}
\label{fg:diagram}
\end{figure}

\begin{figure}
\caption{Phase diagrams derived from the present model with 
temperature-dependent $b_0$ defined 
in eq.(\protect\ref{eq:b0change}). (with energy unit $|\delta \epsilon^{\rm T}|$) 
a)For a poorer solvent 
$k_B\theta=1.0|\delta\epsilon^{\rm T}|$ and 
b)a better solvent $k_B\theta=0.5|\delta\epsilon^{\rm T}|$. Notation and values of parameters 
used are the same as 
those of Fig.\protect\ref{fg:diagram} in addition to the random-coil phase, 
which we denote by $\rm C$.}
\label{fg:diagram2}
\end{figure}

\parskip 20pt \par\noindent

\begin{thebibliography}{99}
\bibitem{Kuhn}
In this paper, we use an effective Gaussian chain that does not distinguish 
the Kuhn length from the monomer-monomer distance.
Since protein is one of the more flexible polymers, this simplification may not be too
crude. See for example \cite{Grosberg94} for more detail.
\bibitem{volume}
We note that this may be too crude especially near the $\theta$ temperature 
where the density fluctuations are large. For this possibility, see for 
example, S. S. Plotkin, J. Wang, and P. G. Wolynes, XXXX. 
Here we do not concentrate on the behavior near the $\theta$ temperature but
use this approximation which is relevant when motion occurs within the ensemble of 
strongly collapsed states.
\bibitem{ignore}
Here, we modified the last term in eq.(\ref{eq:Ffolded}) based on the 
physical interpretation; as mentioned before, the last term is expected to cancel 
out the effective attraction due to the randomness in the first term of the same 
expression. Since we should not take the numerical coefficient too seriously in this 
level of calculation, we ignore the factor $2^{-3/2}$ so that the cancellation 
takes place completely.
\bibitem{on_gap}
Although the quantity to be subtracted, $\rho^*vb_0/4$, comes from 
truncation of virial expansion, the gap here is well-defined quantity within the model. 
We also note that what is important is basically a relative change of 
$|\delta\epsilon^{\rm T}|$ but not the value itself. A larger gap corresponds 
to a more designed sequence to the fast folding.
\end{thebibliography}
\end{document}